\documentclass[12pt]{article}
\usepackage[cp866]{inputenc}
\usepackage[T2A]{fontenc}
\usepackage{amsmath,amsfonts}
\usepackage{graphicx}

\begin{document}
\font\bss=cmr12 scaled\magstep 0
\title{The semiclassical tunneling probability
in quantum cosmologies with varying speed of light }
\author{ A.V. Yurov, V.A. Yurov \\
\small  Theoretical Physics Department,\\
\small  Kaliningrad State
University, \\
\small 236041, Al.Nevsky St., 14, Kaliningrad, Russia \\
\small   e-mails: artyom\_yurov@mail.ru; yurov@freemail.ru}
\date {}
\renewcommand{\abstractname}{\small Abstract}
\maketitle
\begin{abstract}
In quantum cosmology the closed universe can spontaneously
nucleate out of the state with no classical space and time. The
semiclassical tunneling nucleation probability can be estimated as
$\emph{P}\sim\exp(-\alpha^2/\Lambda)$ where $\alpha$=const and
$\Lambda$ is the cosmological constant.

In classical cosmology with varying speed of light $c(t)$ (VSL) it
is possible to solve the horizon problem, the flatness problem and
the $\Lambda$-problem if $c=sa^n$ with $s$=const and  $n<-2$. We
show that in VSL quantum cosmology with $n<-2$ the semiclassical
tunneling nucleation probability is
$\emph{P}\sim\exp(-\beta^2\Lambda^k)$ with $\beta$=const and
$k>0$. Thus, the semiclassical tunneling nucleation probability
in VSL quantum cosmology is very different from this one in
quantum cosmology with $c$=const. In particular, this one is
strongly suppressed for large values of $\Lambda$.
\end{abstract}
\thispagestyle{empty}
\medskip
\section{Introduction.}
Among the most interesting alternatives to inflation in physics
nowadays are truly the cosmological models with varying speed of
light (VSL)\cite{1}, \cite{2}~\footnote{In fact, there are many
articles about this matter. But we'd like to restrict ourselves
to consider only those ones which has been used in this work.}.
In simplest case the speed of light $c =c(t)$ varies as some
power of the expansion scale factor: $c(t)=s a^n(t)$, where
constant $s>0$. Summarizing some of the promising positive
features of these models:
\newline
1. It can solve the horizon problem if $n<0$.
\newline
2. It can solve the flatness problem in a radiation-dominated
early universe if $n<-1$.
\newline
3. In case of $n<-2$ the VSL models can solve the
$\Lambda$-problem in a radiation-dominated early universe while
inflation models can't handle it without the aid of the anthropic
principle.

Of course, these VSL models result in some shortcomings and
unusual (unphysical?) features as well \cite{3}:
\newline
1. It is not clear how to solve the isotropy problem.
\newline
2. The quantum wavelengths of massive particle states and the
radii of primordial black holes can grow sufficiently fast to
exceed the scale of the particle horizon.
\newline
3. The entropy problem: Entropy can decrease with increasing time.

Keeping in mind all the above-mentioned problems we'd like,
nevertheless, to consider VSL quantum cosmology.  There are known
the three ways to describe quantum cosmology: the Hartle-Hawking
wave function \cite{4}, the Linde wave function \cite{5}, and the
tunneling wave function \cite{6}. In the last case the universe
can tunnel through the potential barrier to the regime of
unbounded expansion with semiclassical probability
$\emph{P}\sim\exp(-\alpha^2/\Lambda)$. For the universe filled
with radiation this result can be obtained in framework of a
standard WKB wave function for a particle in a potential
$U(a)=c^2a^2(1-\Lambda a^2/3)$, with the small energy $E$, see
Fig.1. We have two Lorentzian regions ($0<a<a'_i$, $a>a_i$) and
one Euclidean region ($a'_i<a<a_i$). The second turning point
$a=a_i$ corresponds to the beginning of our universe. If
$\Lambda=0$ then $U(a)$ has the form of {\bf parabola} and we get
only one Lorentzian region (see Fig.1). In this case, the
universe can start at $a=0$, expand to a maximum radius  and then
recollapse. If $E\to 0$ then the single Lorentzian region is
contract to the point. This, of course, comes to an agreement with
the tunnelling nucleation probability: $\emph{P}\to 0$ as
$\Lambda\to 0$.

In this article, however, we'll show that in quantum cosmological
VSL models the situation can be opposite, viz: the probability to
find the finite universe short after it's tunneling through the
potential barrier is
$\emph{P}\sim\exp(-\beta(n)\Lambda^{\alpha(n)})$ with
$\alpha(n)>0$ and $\beta(n)>0$ when $n<-2$ or for $-1<n<-2/3$.
After the tunneling one get the finite universe with "initial"
value of scale factor $a_i\sim\Lambda^{-1/2}$, so the probability
to find the universe with large value of $\Lambda$ and small
value of $a_i$ is strongly suppressed. The reason of this is that,
for the case $\Lambda\to 0$, the potential $U(a)$ is transformed
into the {\bf hyperbola} which is located under the abscissa axis
and thus such a universe can start at $a\sim 0$ the regime of
unbounded expansion (see Fig.2). As a result, we get the single
Lorentzian region which is not contract to the point at $E\to 0$.

\section {Albrecht-Magueijo-Barrow VSL model}
Lets start with the Friedmann and Raychaudhuri system of equations
with $k=+1$ (we assume that $G$=const):
\begin{equation}
\begin{array}{cc}
\displaystyle{\frac{\ddot a}{a}=-\frac{4\pi
G}{3}\left(\rho+\frac{3 p}{c^2}\right)+\frac{\Lambda c^2}{3},}
\\
\\
\displaystyle{\left(\frac{\dot a}{a}\right)^2=\frac{8\pi
G\rho}{3}-\left(\frac{c}{a}\right)^2+\frac{\Lambda c^2}{3},}
\\
\\
\displaystyle{c=c_0\left(\frac{a}{a_0}\right)^n=sa^n,\qquad
p=wc^2\rho,}
\end{array}
\label{frid}
\end{equation}
where $a=a(t)$ is the expansion scale factor of the Friedmann
metric, $p$ is the fluid pressure, $\rho$ is the fluid density,
$k$ is the curvature parameter, $\Lambda$ is the cosmological
constant, $c_0$ is the modern value of speed of light ($3\times
10^{10}$ sm/sec) and $a_0$ is the modern value of the scale
factor ($a_0\sim 10^{28}$ sm).

Using (\ref{frid}) one get
\begin{equation}
\dot\rho=-\frac{3\dot
a}{a}\left(\rho+\frac{p}{c^2}\right)+\frac{{\dot
c}c(3-a^2\Lambda)}{4\pi G a^2}. \label{rt}
\end{equation}
Choosing $w=1/3$ one can solve (\ref{rt}) to receive
\begin{equation}
\rho=\frac{M}{a^4}+\frac{3s^2na^{2(n-1)}}{8\pi G(
n+1)}-\frac{s^2n\Lambda a^{2n}}{8\pi G(n+2)}, \label{rho}
\end{equation}
where $M>0$ is a constant characterizing the amount of radiation.
It is clear from the (\ref{rho}) that the flatness problem can be
solved in a radiation-dominated early universe by an interval of
VSL evolution if $n < -1$, whereas the problem of $\Lambda$-term
can be solved only if $n<-2$.  The evolution equation for the
scale factor $a$ (the second equation in system (\ref{frid})) can
be written as
\begin{equation}
p^2+U(a)=E,\qquad E\ge 0, \label{equation}
\end{equation}
where $p=-a\dot a$ is the momentum conjugate to $a$, $E=8\pi G
M/3$ and
\begin{equation}
U(a)=\frac{s^2a^{2n+2}}{n+1}-\frac{2s^2\Lambda a^{2n+4}}{3(n+2)}.
\label{U}
\end{equation}
The potential (\ref{U}) has one maximum at
$a=a_e=\sqrt{3/(2\Lambda)}$ such that
\begin{equation}
U_e\equiv
U(a_e)=\frac{s^23^{n+1}}{2^{n+1}\Lambda^{n+1}(n+1)(n+2)},
\label{Ue}
\end{equation}
so $U_e>0$ if (i) $n<-2$ or (ii) $n>-1$. The first case allows us
to solve the flatness and "Lambda" problems. The surplus dividend
of the model  is the presence of finite time region during which
universe has accelerated expansion (see Appendix).

\section{The semiclassical tunneling probability
in VSL models with $n<-2$: the case $E\ll U_e$}

One can choose $n=-2-m$ with $m>0$. Such a substitution gives us
the potential (\ref{U}) in the form
\begin{equation}
U_m(a)=\frac{s^2}{a^{2(m+1)}}\left(\frac{2\Lambda
a^2}{3m}-\frac{1}{m+1}\right). \label{Um}
\end{equation}

The equation (\ref{equation}) is quite similar to equation for the
particle of energy $E$ that is moving in potential (\ref{Um}),
hence the universe in quantum cosmology can start at $a\sim 0$,
expand to a maximum radius $a'_i$ and then tunnel through the
potential barrier to the regime of unbounded expansion with
"initial" value $a=a_i$, see Fig.2. The semiclassical tunneling
probability can be estimated as
\begin{equation}
\emph{P}\sim\exp\left(-2\int_{a'_i}^{a_i} {\mid {\tilde p(a)}\mid}
da\right), \label{P}
\end{equation}
with
$$
{\mid {\tilde p(a)}\mid}=\frac{c^2(t)}{8\pi G\hbar}\mid
p(a)\mid,\qquad \mid p(a)\mid=\sqrt{U_m(a)-E},\qquad E\le U_e.
$$
It is  convenient to write $E=U_e\sin^2\theta$, with
$0<\theta<\pi/2$.

For the case $E\ll U_e$ one can choose
\begin{equation}
\displaystyle{ a'_i\sim
a_1=\sqrt{\frac{3m}{2(m+1)\Lambda}},\qquad a_i\sim
\sqrt{\frac{3}{2\Lambda}}\left(\frac{\sqrt{m+1}}{\sin\theta}\right)^{1/m}
,} \label{ai-ai}
\end{equation}
 and evaluate the integral (\ref{P}) as
\begin{equation}
\displaystyle{
\emph{P}\sim\exp\left(\frac{-s^3\Lambda^{2+3m/2}I_m(\theta)}{4\pi
G\hbar}\right),} \label{P1}
\end{equation}
where
\begin{equation}
I_m(\theta)=\int_{z'_i(\theta)}^{z_i(\theta)} dz
z^{-5-3m}\sqrt{\frac{2z^2}{3m}-\frac{1}{m+1}}, \label{Im}
\end{equation}
with
$$
 z'_i(\theta)=\sqrt{\frac{3m}{2(m+1)}},\qquad
z_i(\theta)=\sqrt{1.5}\left(\frac{(m+1)^{1/2}}{\sin\theta}\right)^{1/m}.
$$
The integral (\ref{Im}) can be calculated for the $m\in Z$. For
example
$$
I_1(\theta)=\frac{\sqrt{3}}{17010}\left(3+\cos^2\theta\right)\left(191-78
\cos^2\theta+15 \cos^4\theta\right)\sqrt{6+2 \cos^2\theta}\sim
0.148+O(\theta^6),
$$
$$
I_2\sim 0.025,\qquad I_3\sim 0.007,\qquad I_4\sim 0.002,
$$
and so on. One can further show that $I_m(\theta)>0$ at
$0<\theta\ll 1$. Thus, it is easy to see from (\ref{P1}) that the
semiclassical tunneling probability $\emph{P}\to 0$ for large
values of $\Lambda>0$ and $\emph{P}\to 1$ at $\Lambda\to 0$.

Note, that the case $c$=const can be obtained by substitution
$m=-2$ into the (\ref{P1}). Not surprisingly, this case will get
us the well known result $\emph{P}\sim\exp(-1/\Lambda)$ (see
\cite{7}).

In quantum cosmology the closed universe can spontaneously
nucleate out of a state with no classical space and time. It mean
that we must choose
\begin{equation}
(a'_i)^2\le a^2_{Pl}(t)=\frac{\hbar G}{c^3(t)}.
\label{nerav}
\end{equation}
Substituting (\ref{ai-ai}) into (\ref{nerav}) one  get the
inequality,
\begin{equation}
\Lambda\le \Lambda_m=\frac{3m}{2(m+1)}\left(\frac{\hbar
G}{c^3_0a_0^{3(m+2)}}\right)^{2/(4+3m)}, \label{ineq}
\end{equation}
where $c_0$ is the modern value of speed of light and $a_0$ the
modern value of scale factor (or the size of visible universe:
$a_0\sim 10^{28}$ sm). For example
$$
\Lambda_1=0.14\times 10^{-90}\,\,\,{\rm sm}^{-2},\,\,\,
\Lambda_2=0.48\times 10^{-80}\,\,\,{\rm sm}^{-2},\,\,\,
\Lambda_3=0.22\times 10^{-74}\,\,\,{\rm sm}^{-2},\,\,\,
\Lambda_{10}=10^{-63}\,\,\,{\rm sm}^{-2}.
$$
For the case $m\ll 1$ we get
$$
\frac{4\Lambda^2c^3_0a^6_0}{9\hbar G}\le m^2\ll 1,
$$
so $\Lambda\ll 0.16\times 10^{-116}$ ${\rm sm}^{-2}$.

It is convenient to introduce the cosmological parameter
$\Omega_0=\mid \rho_{_{\Lambda}}(t_0)\mid /\rho_c(t_0)$, where
$\rho_{_{\Lambda}}(t_0)$ is the contribution of $\Lambda$ to
modern value of density and $\rho_c(t_0)=10^{-29}$ ${\rm
{gramme/sm^{3}}}$ is the critical density. One get
$$
\Omega_0=\left|
\frac{\rho_{_{\Lambda}}(t_0)}{\rho_c(t_0)}\right|=\frac{(m+2)c^2_0\Lambda}{8\pi
G m\rho_c(t_0)}\le \Omega_m=\frac{(m+2)c^2_0\Lambda_m}{8\pi G
m\rho_c(t_0)}.
$$
So
$$
\Omega_1=0.2\times 10^{-34},\qquad \Omega_2=0.5\times
10^{-24},\qquad \Omega_{10}=0.6\times 10^{-7},\qquad
\Omega_{20}=1.3\times 10^{-4},
$$
and so on. Hence, in framework of this model, the whole
contribution of $\Lambda$ in modern era is a quite negligible
quantity (And that's just how it should be.).

The obtained values of $\Lambda$ are seemingly unnatural. Of
course, we can assume that $a'_i>a_{Pl}$, in spite of  the
(\ref{nerav}). In this case, we receive the following picture:
there is the pre-Big-Bang universe which can start its evolution
at $a\sim a_{Pl}$, expand to a maximum radius $a'_i$ and then
tunnel through the potential barrier to the regime of unbounded
expansion i.e. turn into the universe we see nowadays. The
probability of this process will be large for small value of
$\Lambda$. But the $\Lambda$-term in our universe will be the
same as in pre-Big-Bang universe. Thus, the above scenario
doesn't explain the reason why the $\Lambda$ was so small in
pre-Big-Bang universe and how the pre-Big-Bang universe was born.
Therefore, if we want to describe the quantum nucleation of
universe we must use the (\ref{nerav}). As we have seen, the
requirement for this is the verity of (\ref{ineq}), i.e. the
smallness of $\Lambda$ and it is highly uncommon that for such
values of $\Lambda$, the probability of quantum nucleation, in
fact, will be so large.

\section{The semiclassical tunneling probability with $n<2$ and $n>-1$}

In the case of general position the semiclassical tunneling
probability with $n=-2-m$ has the form
\begin{equation}
\emph{P}_m\sim \exp\left(-\frac{s^3\Lambda^{(3m+4)/2}}{4\pi G\hbar
3^{(m+1)/2}\sqrt{m(m+1)}}\int_{z'_i}^{z_i}
\frac{dz}{z^{3m+5}}\sqrt{F_m(z,\theta)}\right), \label{Pm}
\end{equation}
where
\begin{equation}
F_m(z,\theta)=-2^{m+1}\sin^2\theta\, z^{2(m+1)}+2\times 3^m (m+1)
z^2-m 3^{m+1}, \label{Fm}
\end{equation}
$z$ is dimensionless quantity and  $z'_i$, $z_i$ are the turning
points, i.e. two real positive solutions of the equation
$F_m(z,\theta)=0$ for the given $\theta$ (it is easy to see that
the equation $F_m(z,\theta)=0$ does have two such solutions at
$0<\theta<\pi/2$).

If $m$ is the natural number then the expression (\ref{Pm}) has a
more simple form. For example
\begin{equation}
\emph{P}_1\sim\exp\left(-\frac{s^3 \Lambda^{7/2}\sin\theta }{6\pi
G\hbar\sqrt{2}}\int_{z'_i}^{z_i}\frac{dz}{z^8}\sqrt{(z^2-{z'_i}^2)(z_i^2-z^2)}\right),
\label{P1}
\end{equation}
with
$$
z'_i=\frac{\sqrt{3}}{2\cos(\theta/2)},\qquad
z'_i=\frac{\sqrt{3}}{2\sin(\theta/2)}.
$$
This expression can be calculated exactly:
\begin{equation}
\emph{P}_1\sim\exp\left(-\frac{s^3\Lambda^{7/2}\sin\theta
J(\theta)}{6\sqrt{2}\pi G\hbar}\right), \label{m=1}
\end{equation}
with
$$
J(\theta)=\frac{1}{105}\left(\frac{2\sin(\theta/2)}{\sqrt{3}}\right)^5
\left[\frac{8\lambda^4-13\lambda^2+8}{\cos^2(\theta/2)}
\Pi\left(\mu^2;\frac{\pi}{2}{\backslash}\arcsin\mu\right)-
2\left(2\lambda^4-\lambda^2+2\right) \textrm{K}(\mu^2)\right],
$$
where $\mu^2=\cos\theta/\cos^2(\theta/2)$,
$\lambda=\cot(\theta/2)$, $\Pi$ and $\textrm{K}$ are complete
elliptic integral of the first and the third kinds correspondingly
\cite{8}.

Similarly,
\begin{equation}
\emph{P}_2\sim\exp\left(-\frac{s^3 \Lambda^5\sin\theta }{18\pi
G\hbar}\int_{z'_i}^{z_i}\frac{dz}{z^{11}}\sqrt{(z^2+z_1^2)(z^2-{z'_i}^2)(z_i^2-z^2)}\right),
\label{P2}
\end{equation}
where
$$
z_1=\sqrt{\frac{3}{\sin\theta}\cos\left(\frac{\theta}{3}-\frac{\pi}{6}\right)},\qquad
z'_i=\sqrt{\frac{3}{\sin\theta}\sin\frac{\theta}{3}},\qquad
z_i=\sqrt{\frac{3}{\sin\theta}\cos\left(\frac{\theta}{3}+\frac{\pi}{6}\right)},
$$
and so on.

Therefore the probability to obtain (via quantum tunneling
through the potential barrier) the universe in the regime of
unbounded expansion is strongly suppressed for large values of
$\Lambda$ and small values of the initial scale factor $a_i
=\sqrt{3}/(2\sin(\theta/2)\sqrt{\Lambda})$. In other words,
overwhelming majority of universes which are nascent via quantum
tunneling through the potential barrier (\ref{U}) have large
initial scale factor and small value of $\Lambda$. Furthermore,
it's clear that this case permits to obtain the analog of the
(\ref{ineq}) too.


Now, let us consider the case (ii), when $n>-1$. The "quantum
potential" has the form
\begin{equation}
U(a)=s^2a^{2m}\left(\frac{1}{m}-\frac{2\Lambda
a^2}{3(m+1)}\right), \label{UU}
\end{equation}
where $m=n+1>0$, see Fig. 3. The points of intersection with the
abscissa axis $a$ are $a_0=0$ and $a_1=\sqrt{3(m+1)/2\Lambda m}$.
Choosing $E=0$ in equation (\ref{equation}) and substituting
(\ref{UU}) into the (\ref{P}) we get
\begin{equation}
\emph{P}\sim\exp\left(-\frac{s^3\Lambda^{(1-3m)/2}}{4\pi
G\hbar}\int_0^{z_1}
z^{2m-2}\sqrt{\frac{1}{m}-\frac{2z^2}{3(m+1)}}\,dz\right),
\label{PP}
\end{equation}
with $z_1=\sqrt{3(m+1)/2m}$~\footnote{The starting value $z=0$
means that the Universe tunneled from "nothing" to a closed
universe of a finite radius $a_1=z_1/\sqrt{\Lambda}$.}. Thus, we
have the same effect as if $0<m<1/3$.

The density $\rho$ is
\begin{equation}
\rho=\frac{M}{a^4}-\frac{3(1-m)s^2}{8\pi Gm
a^{4-2m}}+\frac{s^2(1-m)\Lambda}{8\pi G(1+m)a^{2-2m}}. \label{rho}
\end{equation}
The second term is the curvature term, whereas the third term is
the $\Lambda$-term. It is easy to see that at large $a$ the first
term falls off faster than the curvature term which in turn falls
off faster than the $\Lambda$-term. At the beginning, when $t=t_1$
and $a(t_1)=a_1$ we have
$$
\rho(t_1)=-\rho_{\Lambda}(t_1)=-\frac{s^2(1-m)\Lambda^{2-m}m^{1-m}}{2^{2+m}3^{1-m}\pi G(1+m)^{2-m}},
$$
therefore
$$
\frac{{\ddot{a}}(t_1)}{a(t_1)}=-\frac{4\pi
G}{3}\left(\rho(t_1)+3\frac{p(t_1)}{c^2(t_1)}\right)+\frac{\Lambda
c^2(t_1)}{3}= \frac{1}{3}\left(8\pi
G\rho_{\Lambda}(t_1)+\frac{\Lambda s^2}{a_1^{2(1-m)}}\right)>0.
$$
In other words, we have accelerated expansion at the beginning. As times goes by, the $\Lambda$-term will be
dominating term so
$$
a(t)\sim \left(s\sqrt{\frac{\Lambda(1-m)^3}{1+m}} t\right)^{1/(1-m)}\sim t^{\kappa},
$$
with $\kappa=1/(1-m)\in (1,\,1.5)$, at $t\to \infty$. Therefore
for $t\to \infty$ we have $\ddot a<0$. It means that accelerated
expansion continue during a finite time.
\section{Peculiar cases with $n=-1$ and $n=-2$}
Now, lets consider the cases of $n=-1$ and $n=-2$. The formula
(\ref{Pm}) is not valid in these cases ($m=-1$ and $m=0$) so we
shall consider these models separately.

If $n=-1$ ($m=-1$) then
$$
\rho=\frac{M}{a^4}+\frac{\Lambda s^2}{8\pi Ga^2}-\frac{3s^2}{4\pi
G a^4}\log\frac{a}{a_*},
$$
therefore
\begin{equation}
U(a)=s^2\left(2\log\left(\frac{a}{a_*}\right)-\frac{2a^2\Lambda}{3}+1\right),
\label{U-1}
\end{equation}
where $a_*$ are constant and $[a_*]$=sm. The potential
(\ref{U-1}) has one maximum at $a=a_e=\sqrt{3/(2\Lambda)}$ such
that $U_e=U(a_e)=2s^2\log(a_e/a_*)$, so if $a_e>a_*$ then $U_e>0$
(see Fig. 4) . We choose $a_*=\Lambda^{-1/2}$. This gives us
$U_e=0.41 s^2>0$. For the case $E\ll U_e$ the semiclassical
tunneling nucleation probability is
\begin{equation}
\emph{P}_{_{-1}}\sim\exp\left(-\frac{s^3\sqrt{\Lambda}}{4\pi
G\hbar}\int_{z'_i}^{z_i}\frac{dz}{z^2}\sqrt{\log
z^2-\frac{2z^2}{3}+1}\right)\sim
\exp\left(-\frac{s^3\sqrt{\Lambda}}{10\pi G\hbar}\right),
\label{Pr1}
\end{equation}
where the turning points are $z'_i=0.721$, $z_i=1.812$. As we can
see from the (\ref{Pr1}), when $n=-1$ we receive the
aforementioned effect again.

If $n=-2$ ($m=0$) then
$$
\rho=\frac{M}{a^4}+\frac{s^2\Lambda}{2\pi
Ga^4}\log\left(\frac{a}{a_*}\right)+\frac{3s^2}{4\pi Ga^6}.
$$
We choose $a_*=1/(\alpha\sqrt{\Lambda})$, where $\alpha$ is a
dimensionless quantity. Thus
\begin{equation}
U(a)=-s^2\left(\frac{1}{a^2}+\frac{4\Lambda}{3}\log\left(\alpha
a\sqrt{\Lambda}\right)+\frac{\Lambda}{3}\right). \label{U-2}
\end{equation}
The maximum of potential (\ref{U-2}) is located at the same point
$a_e$ and
$$
U_e=-\frac{s^2\Lambda}{3}\left(3+\log\left(\frac{9\alpha^4}{4}\right)\right).
$$
Therefore, $U_e>0$ if $\alpha<2{\rm e}^{-3/4}/\sqrt{6}\sim
0.386$. Choosing $\alpha=0.286$ and $E\ll U_e$ gets us the turning
points $z'_i\sim 0.77$ and $z_i\sim 2.391$. The potential is
pictured on the Fig.5.

At last, the semiclassical tunneling nucleation probability is
$$
\emph{P}_{0}\sim\exp\left(-\frac{s^3\Lambda^2}{4\pi
G\hbar}\int_{z'_i}^{z_i}\frac{dz}{z^4}\sqrt{-\frac{1}{z^2}-\frac{4}{3}\log(\alpha
z)-\frac{1}{3} }\right)\sim \exp\left(-\frac{0.084
s^3\Lambda^2}{\pi G\hbar}\right).
$$


\section{Conclusion}
As we have shown, the semiclassical tunneling nucleation
probability in VSL quantum cosmology is quit different from the
one in quantum cosmology with $c$=const. In the first case this
probability  is strongly suppressed for large values of $\Lambda$
whereas in the second case it is strongly suppressed for small
values of $\Lambda$. Nevertheless, we can't really say that VSL
quantum cosmology results in solution of the $\Lambda$-mystery.
The problem is the validity the WKB wave function. And what is
more, we have omitted all preexponential factors which can be
essential ones near the turning points.

But, all in all, the difference between $\emph{P}$ in VSL and
usual quantum cosmology seems very interesting and very
significant.

$$
{}
$$
 ACKNOWLEDGMENTS
\newline
\newline
After finishing this work, we learned that T.Harko, H.Q.Lu,
M.K.Mak and K.S.Cheng \cite{Harko}, have independently considered
the VSL tunneling probability in both Vilenkin and Hartle-Hawking
approaches.  The interesting conclusion of their work is that at
zero scale factor the classical singularity is no longer isolated
from the Universe by the quantum potential but instead classical
evolution can start from arbitrarily small size. In contrast to
\cite{Harko}, we attract attention to the fact that the
semiclassical tunneling nucleation probability in VSL quantum
cosmology is strongly suppressed for large values of $\Lambda$
and small values of the initial scale factor $a_i\sim
1\sqrt{\Lambda}$.

We'd like to thank Professor Harko for useful information about
the article \cite{Harko}.
\section*{Appendix} Lets take $n=-2-m$, for the $m\ge 0$. Substituting
(\ref{rho}) into the first equation of system (\ref{frid}) yields
$$
{\ddot
a}=\frac{1}{a^3}\left[-E+\frac{s^2}{a^{2m}}\left(-\frac{m+2}{(m+1)a^2}+\frac{2(m+1)\Lambda}{3m}\right)\right].
\eqno(A1)
$$

Thus we have the following situation:
\newline
1. If
$$
0<a^2\ll \frac{3m(m+2)}{2\Lambda(m+1)^2},
$$
then the curvature term is the dominating one and ${\ddot a}<0$.
\newline
2. If
$$
\frac{3m(m+2)}{2\Lambda(m+1)^2}\ll a^2\ll {\tilde a}^2\equiv
\left(\frac{2s^2(m+1)\Lambda}{3mE}\right)^{1/m}, \eqno(A2)
$$
then the dominating term is $\Lambda$-term and $\ddot a>0$ during
this time.
\newline
3. If
$$
a^2\gg \left(\frac{2s^2(m+1)\Lambda}{3mE}\right)^{1/m},
$$
then the radiation term is the dominating one and ${\ddot a}<0$.

There are two way to interpret the region (A2). The first way is
to conclude that we have cosmological inflation in early
universe. This is possible when $0<m\ll 1$. In this case we can
evaluate the number of e-foldings  $\Delta N$ during the region
(A2) as
$$
\log m\sim -2 m\Delta N,\qquad  \Lambda\gg \frac{3Em}{2s^2}\sim
\frac{3Em}{2c_0^2a_0^4}.
\eqno(A3)
$$
If $\Delta N\sim 60$ then $m\sim 0.029$; if $\Delta N\sim 100$
then $m\sim 0.0197$. To evaluate $E$ one can use the well-known
expression for the Friedmann integrals \cite{Cher},
$$
A(w)=\left[\left(\frac{1+3w}{2}\right)^2 E\right]^{1/(1+3w)}.
$$
Since $A(1/3)=3\times 10^{36}$ ${\rm sm}^2/{\rm sec}$, we get
$E=0.9\times 10^{73}$ ${\rm sm}^4/{{\rm sec}^2}$. The
substitution of $A(1/3)$ into the (A3) results in
$$
\Lambda\gg 0.435\times 10^{-61}\,\,{\rm sm}^{-2},\qquad \Lambda\gg
0.296\times 10^{-61}\,\,{\rm sm}^{-2},
$$
for the $\Delta N=60$ and $\Delta N=100$.

But do we really need inflation in the VSL models? The question
is not quite clear. On the one hand, VSL models can solve
fundamental cosmological problems (horizon and  flatness
problems) without inflation - and what is more, these models can
solve $\Lambda$-problem whereas inflations can't do it without the
anthropic principle. On the other hand, the simplest case of VSL
cosmological models, which is the subject of this article, is
facing with the isotropy problem \cite{3}. But, as we have seen,
VSL model results in inflation with exit naturally so it will be
incorrectly to oppose VSL models and the inflation.

Another way to interpret the region (A2) is to identify this
region with modern acceleration of universe. This is possible if
$m$ is sufficiently large. Let us make a crude guess. According to
modern observations we can write ${\ddot a_0}/a_0=5.6\pi
G\rho_c/3$ where $\rho_c=10^{-29}$ ${\rm {gramme/sm^{3}}}$. If
the modern value of $a_0\sim{\tilde a}$ (see the inequality (A2))
then
$$
\Lambda=\frac{3mE}{3c_0^2(m+1)a^4_0}. \eqno(A4)
$$
From the (A1) we have
$$
\ddot a_0\sim\frac{2(m+1)\Lambda s^2}{3ma^{2m+3}_0}
$$
if the $\Lambda$-term is dominating one.  Substituting (A4) gets
us
$$
\Omega_{_{\Lambda}}=\frac{\Lambda c_0^2}{8\pi G\rho_c}\sim
\frac{0.35 m}{m+1},\qquad a_0\sim{}^4\sqrt{\frac{3E}{5.6\pi G
\rho_c}}.
$$
Thus, if $m\gg 1$ then $\Omega_{_{\Lambda}}\sim 0.35$,
$\Lambda\sim 0.2\times 10^{-56}$ ${\rm sm}^{-2}$, $a_0\sim
10^{27}$ sm. And these values are seems to be quite reasonable.
$$
{}
$$

\vfill
\eject

\end{document}